\documentclass[reprint, superscriptaddress, twocolumn, amsmath, amssymb, aps, prb]{revtex4-2}
\usepackage{graphicx}% Include figure files
\usepackage{dcolumn}% Ali$G_N$ table columns on decimal point
\usepackage{bm}% bold math
\usepackage{placeins}
\usepackage{appendix}
\usepackage{upgreek}

\usepackage{braket} 

\usepackage{verbatim}
\usepackage[usenames,dvipsnames]{xcolor}
 \usepackage{amsmath}
 \usepackage{amsfonts}
 \usepackage{amssymb}
 \usepackage[colorlinks=true,citecolor=blue,linkcolor=red]{hyperref}

 \usepackage{bbold}     % for \mathbb{1} as a nice unit matrix symbol
 \usepackage[makeroom]{cancel}  % crossed terms in tables
 \usepackage{multirow}    % merge cells vertically in tables
 \usepackage[normalem]{ulem}        % strike-through text
 \usepackage{array}
 \usepackage{pdfpages}
\makeatletter
 \AtBeginDocument{\let\LS@rot\@undefined}
 \makeatother

\begin{document}

%\linenumbers

\title{Gate-Compatible Circuit Quantum Electrodynamics in a Three-Dimensional Cavity Architecture}

\author{Zezhou Xia}
\email{equal contribution}
\affiliation{State Key Laboratory of Low Dimensional Quantum Physics, Department of Physics, Tsinghua University, Beijing 100084, China}

\author{Jierong Huo}
\email{equal contribution}
\affiliation{State Key Laboratory of Low Dimensional Quantum Physics, Department of Physics, Tsinghua University, Beijing 100084, China}

\author{Zonglin Li}
\email{equal contribution}
\affiliation{State Key Laboratory of Low Dimensional Quantum Physics, Department of Physics, Tsinghua University, Beijing 100084, China}

\author{Jianghua Ying}
\email{equal contribution}
\affiliation{Yangtze Delta Region Industrial Innovation Center of Quantum and Information, Suzhou 215133, China}

\author{Yulong Liu}
\email{equal contribution}
\affiliation{Beijing Academy of Quantum Information Sciences, Beijing 100193, China}

\author{Xin-Yi Tang}
\affiliation{State Key Laboratory of Low Dimensional Quantum Physics, Department of Physics, Tsinghua University, Beijing 100084, China}

\author{Yuqing Wang}
\affiliation{Beijing Academy of Quantum Information Sciences, Beijing 100193, China}

\author{Mo Chen}
\affiliation{Beijing Academy of Quantum Information Sciences, Beijing 100193, China}

\author{Dong Pan}
\affiliation{State Key Laboratory of Superlattices and Microstructures, Institute of Semiconductors, Chinese Academy of Sciences, P. O. Box 912, Beijing 100083, China}

\author{Shan Zhang}
\affiliation{State Key Laboratory of Low Dimensional Quantum Physics, Department of Physics, Tsinghua University, Beijing 100084, China}

\author{Qichun Liu}
\affiliation{Beijing Academy of Quantum Information Sciences, Beijing 100193, China}

\author{Tiefu Li}
\affiliation{School of Integrated Circuits and Frontier Science Center for Quantum Information, Tsinghua University, Beijing 100084, China}
\affiliation{Beijing Academy of Quantum Information Sciences, Beijing 100193, China}

\author{Lin Li}
\affiliation{Beijing Academy of Quantum Information Sciences, Beijing 100193, China}

\author{Ke He}
\affiliation{State Key Laboratory of Low Dimensional Quantum Physics, Department of Physics, Tsinghua University, Beijing 100084, China}
\affiliation{Beijing Academy of Quantum Information Sciences, Beijing 100193, China}
\affiliation{Frontier Science Center for Quantum Information, Beijing 100084, China}
\affiliation{Hefei National Laboratory, Hefei 230088, China}

\author{Jianhua Zhao}
\affiliation{State Key Laboratory of Superlattices and Microstructures, Institute of Semiconductors, Chinese Academy of Sciences, P. O. Box 912, Beijing 100083, China}

\author{Runan Shang}
\affiliation{Beijing Academy of Quantum Information Sciences, Beijing 100193, China}
\affiliation{Hefei National Laboratory, Hefei 230088, China}

\author{Hao Zhang}
\email{hzquantum@mail.tsinghua.edu.cn}
\affiliation{State Key Laboratory of Low Dimensional Quantum Physics, Department of Physics, Tsinghua University, Beijing 100084, China}
\affiliation{Beijing Academy of Quantum Information Sciences, Beijing 100193, China}
\affiliation{Frontier Science Center for Quantum Information, Beijing 100084, China}

%\date{\today}

\begin{abstract}

Semiconductor-based superconducting qubits offer a versatile platform for studying hybrid quantum devices in circuit quantum electrodynamics (cQED) architecture. Most of these cQED experiments utilize coplanar waveguides, where the incorporation of DC gate lines is straightforward. Here,  we present a technique for probing gate-tunable hybrid devices using a three-dimensional (3D) microwave cavity. A recess is machined inside the cavity wall for the placement of devices and gate lines. We validate this design using a hybrid device based on an InAs-Al nanowire Josephson junction. The coupling between the device and the cavity is facilitated by a long superconducting strip, the antenna. The Josephson junction and the antenna together form a gatemon qubit. We further demonstrate the gate-tunable cavity shift and two-tone qubit spectroscopy. This technique could be used to probe various quantum devices and materials in a 3D cQED architecture that requires DC gate voltages.

\end{abstract}

\maketitle

Superconducting circuits based on Josephson junctions play a crucial role in solid-state quantum information processing \cite{Supremacy}. By replacing the insulating barrier (Al$_{2}$O$_{3}$) in the Josephson element with a semiconductor, new types of qubits, such as gatemons, $0$-$\pi$ qubits, and Andreev qubits, can be realized \cite{2015_PRL_gatemon,DiCarlo_gatemon,Gatemon_two, DiCarlo_gatemon_B, 2D_gatemon, Graphene_Oliver, TI_gatemon, Huo_gatemon, Ramon_perspective, CPH_Pi_qubit, PRL_Hays_Andreev, 2021_Devoret_Science, 2023_NP_Andreev}. In addition, circuit quantum electrodynamics (cQED) provides an approach to exploring the fascinating physics of the semiconductor-superconductor hybrids at microwave frequencies \cite{Steele_Graphene, James_Graphemon, 2019_PRX_Scalay, PRL_Charlie_charge_dispersion, PRL_Delft_charge_dispersion, Chalie_gatemon_fullshell, PRA_CNT_gatemon, 2022_PRL_Devoret, 2022_PRL_Levy, PRX_Quantum_Delft}. These hybrid devices are predicted to exhibit exotic phases of matter, including topological superconductivity \cite{Lutchyn2010, Oreg2010}. While transport measurements have been the primary approach to studying these states \cite{Mourik, Deng2016,Gul2018, WangZhaoyu}, proposals based on their microwave responses offer an additional experimental tool that allows for fast control and readout \cite{NC_Majorana_transmon, Box_qubit, PRL_Cavity_Majorana, Bernard_proposal}.  Previous cQED experiments on these hybrid devices were conducted using a two-dimensional (2D) architecture with superconducting coplanar waveguides \cite{2004_Wallraff}. The incorporation of a DC gate line, which is essential for hybrid devices, is simple in the 2D architecture. To ensure compatibility with an in-plane magnetic field, the superconducting film of the waveguide was often designed to be thin with high-density artificial holes for vortex pinning \cite{Nodar_PRApplied, James_PRApplied, Fluxonium_1T, Gatemon_1T}. However, a magnetic field perpendicular to the substrate can still significantly degrade the performance of the resonator \cite{James_PRApplied}. An alternative approach is the use of a three-dimensional (3D) cavity architecture \cite{Yale_3D}. Incorporating hybrid devices with DC gate lines into a 3D cavity presents considerable challenges. Previous attempts have utilized either a superconducting electrode inserted into the cavity or the cavity itself to apply a DC bias \cite{Wallraff_3D, Steele_3D}. These approaches yield very weak electric fields compared to on-chip gate electrodes for hybrid devices. Directly inserting the device chip with on-chip gate electrodes into a 3D cavity can, however, deteriorate the cavity quality substantially \cite{Guo_3D}.

\begin{figure*}[htb]
\includegraphics[width=0.95\textwidth]{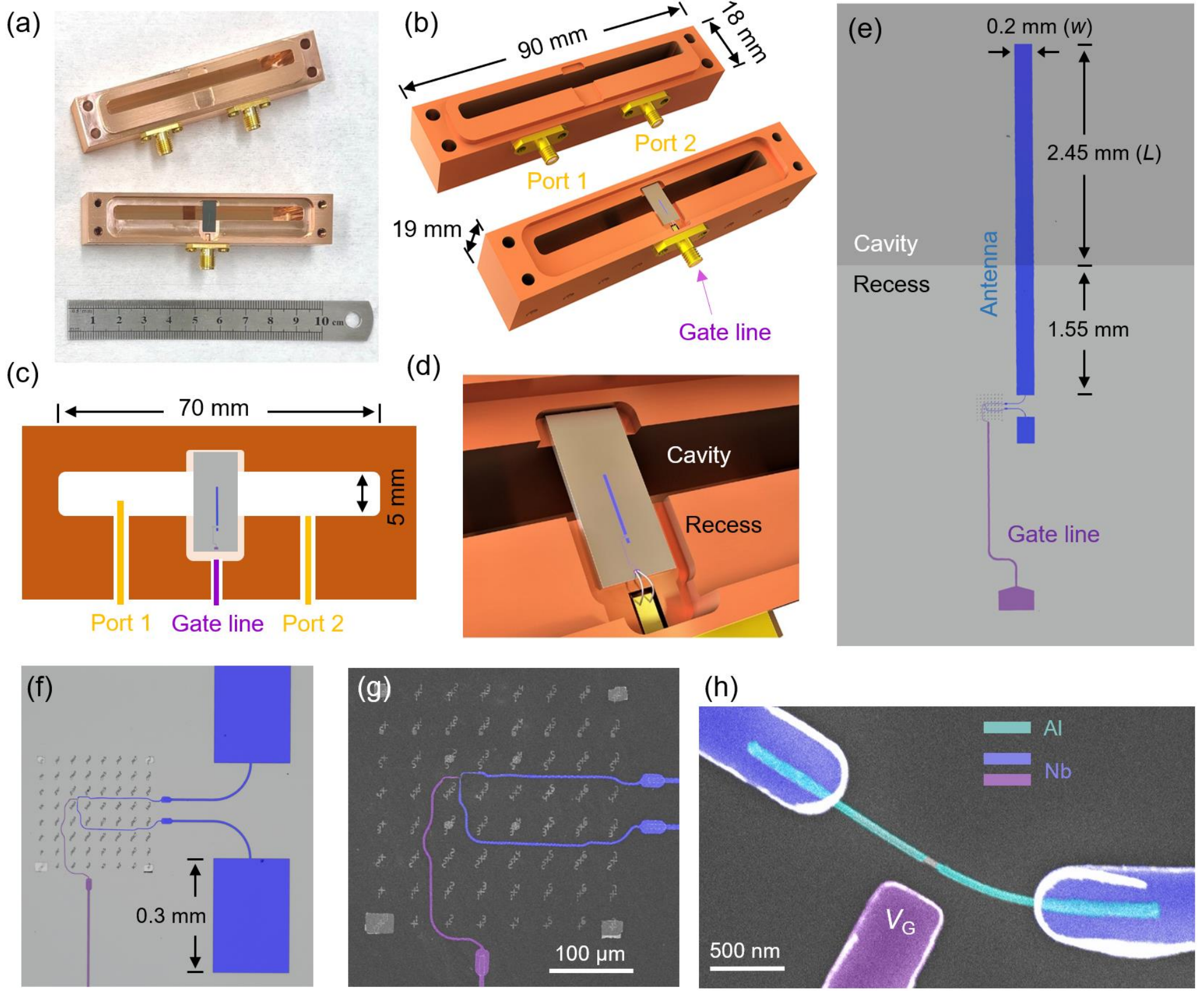}
\centering
\caption{Design of a gate-compatible 3D cQED architecture. (a) Photograph of a 3D copper cavity and a 10-cm-scale ruler. The upper half of the cavity has two commercial SMA connectors (ports 1 and 2) for signal coupling. The lower half has an SMA connector for the gate line. The device chip is in the recess machined within the cavity wall. (b) 3D schematic of the design (not in scale). (c) 2D schematic of the cavity mid-plane. (d) An enlargement of the recess and the device chip (also a schematic). (e) Optical image of the device chip. The dark gray region is false-colored to highlight the cavity region, while the light gray is the part inside the recess. (f) An enlargement of the chip on the device part. (g-h) Device SEMs (false colored).  }
\label{fig1}
\end{figure*}

In this report, we present a 3D cavity architecture that is compatible with a DC gate electrode for probing hybrid devices. For resilience to magnetic fields, we utilized a copper cavity  \cite{Copper_Transmon_Ando, 2019_APL_Copper_qubit, Majumder}. The ohmic dissipation caused by copper should not be an issue, given that decent coherence times ($\sim$ 0.1 ms) have been reported in copper-cavity-based superconducting qubits \cite{Regetti_Copper}. Moreover, the strong thermal anchoring of copper helps in cooling the temperature of the cavity photons and the device chip. The architecture involves machining a recess by ``digging a small room'' on a sidewall of the cavity. The hybrid device, an InAs-Al nanowire Josephson junction, is placed inside this recess. This spatial separation between the device and the cavity can mitigate their direct coupling and minimize unwanted loss. A long superconducting strip, termed the antenna, couples the device to the cavity. The antenna and the InAs-Al nanowire together form a gatemon qubit. We validated this design by demonstrating a gate-tunable shift of the cavity resonance and the qubit spectroscopy. Our technique enables the probing of hybrid superconductor-semiconductor devices in a 3D cQED architecture where gate voltages and a magnetic field are desired.

Figure 1(a) shows a photograph of the cavity-device architecture, with the schematics illustrated in Figs. 1(b-d). The major modification, in comparison to traditional 3D cavities, is a small recess machined on the cavity wall, see Fig. 1(d) for an enlarged view. The height of this recess is larger than the thickness of the device chip, ensuring that the chip can be inserted appropriately, see Fig. S1 in the Supporting Information (SI) for its cross-sectional schematic. Half of the device chip lies within the cavity, while the other half is inserted into the recess. Two SubMiniature version A (SMA) pins (labeled ``Port 1'' and ``Port 2'') penetrate the cavity, connecting the cavity modes to the measurement circuitry. In this work, we measured the reflection coefficient using only port 1, keeping the pin of port 2 grounded and not inserted into the cavity. A DC gate line connects to a third SMA pin, which is bonded onto the device chip. The device and the on-chip gate line are located inside the recess, spatially separated from the cavity. This separation helps to avoid direct coupling between the device region and cavity modes, minimizing unwanted dissipation/loss caused by the gate line and debris/residue from the device fabrication process. As shown in Fig. 1(c), the cavity size is as follows: 70 mm in length, 5 mm in width and 30 mm in height (not drawn). These dimensions yield a resonance frequency of approximately 5 GHz for the TE101 eigenmode.

Figure 1(e) displays an optical image of the lower part of the device chip. Due to space limitations, the upper part of the chip, which extends all the way to the upper sidewall of the cavity, is not shown. The chip substrate is high-resistivity silicon. The long strip, false-colored blue, is a 100-nm thick Nb superconducting film that serves as the antenna for coupling the device and the cavity modes. The width ($w$) of the antenna  is $\sim$ 0.2 mm. One end of the antenna is inserted into the cavity with a length of $L$ $\sim$ 2.45 mm. The other end of the antenna is located inside the recess and is connected to one electrode of an InAs-Al device. Figures 1(f-h) show the optical image and scanning electron micrographs (SEMs). The InAs-Al wire was grown via molecular beam epitaxy \cite{PanCPL}. Quantized zero-bias conductance peaks and peak-to-dip transitions have been reported in these hybrid nanowires as possible signatures of Majorana or quasi-Majorana zero modes \cite{WangZhaoyu, Song2022}. For this experiment, a small segment of the Al shell was etched to form a Josephson junction. The two contacting electrodes are Ti/Nb (1 nm/100 nm) with one connected to the antenna and the other connected to a shorter Nb strip (length 0.3 mm, see Fig. 1(f)). The InAs-Al Josephson junction, its electrodes, and the antenna together constitute a superconducting transmon qubit \cite{Koch_PRA}. As the Josephson coupling $E_J$ can be tuned by a side gate, this type of qubit is also referred to as a gatemon \cite{2015_PRL_gatemon}. The side gate was created in the same lithography step as the contacts. Further details on the device fabrication can be found in SI. 

\begin{figure}[ht]
\includegraphics[width=\columnwidth]{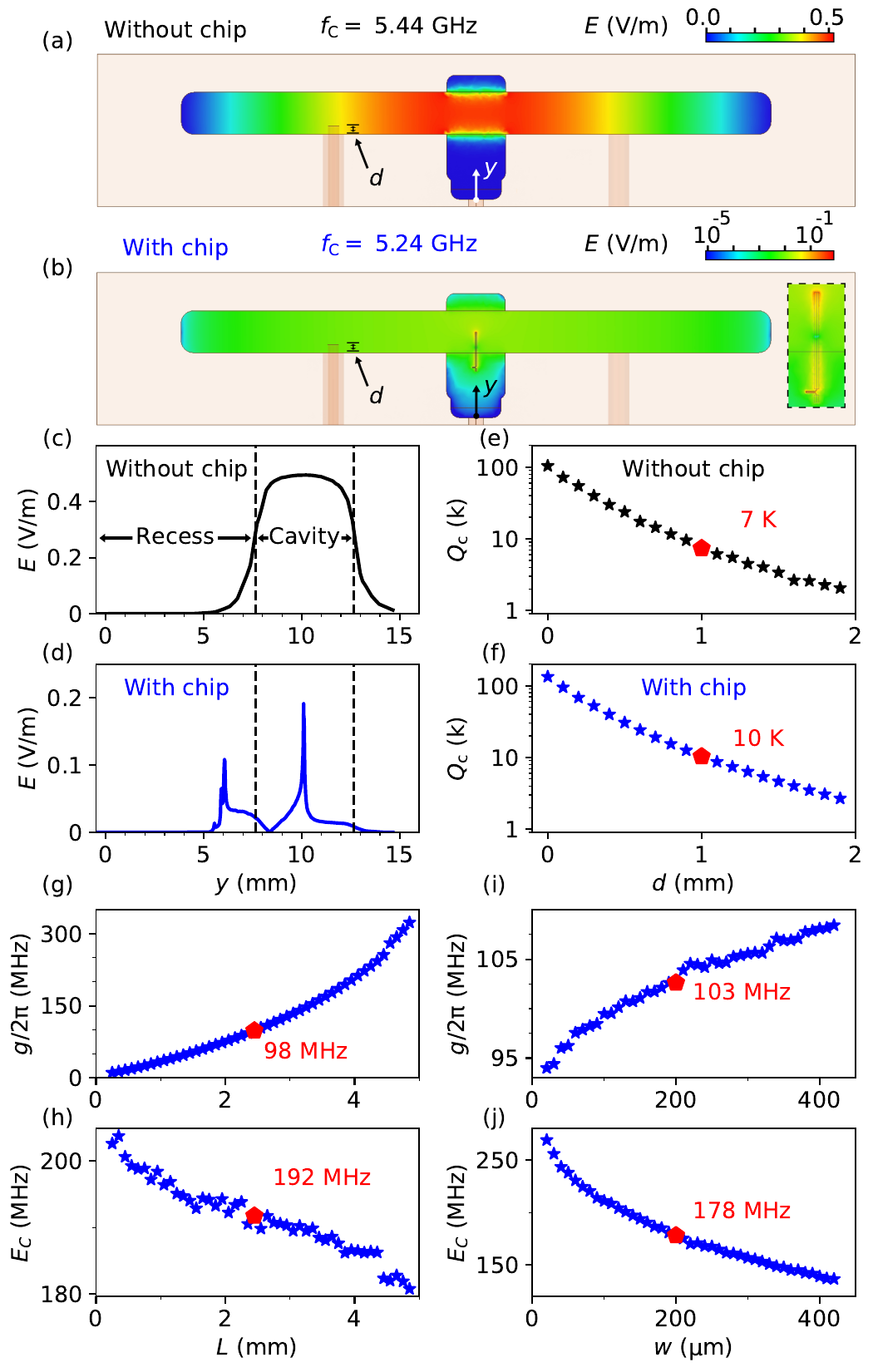}
\centering
\caption{Finite element simulation. (a-b) Electric field ($E$) distribution of the TE101 mode in the cavity without ((a)) and with ((b)) a device chip.  The inset (dashed box) is an enlargement of the antenna region. (c-d) $E$ distribution along the $y$-axis for (a-b), respectively. (e-f) Simulated coupling quality factor $Q_c$ as a function of the penetration length ($d$) of the SMA pin without ((e)) and with ((f)) the device chip. (g) Qubit-cavity coupling strength $g$ as a function of the antenna penetrating length ($L$). (h) Charging energy $E_C$ as a function of $L$. (i-j) $g$ and $E_C$ as a function of the antenna width ($w$). The red dots in (e-j) correspond to the parameters of devices A, B and C. }
\label{fig2}
\end{figure}

To assess the feasibility of this 3D cQED architecture, we performed a finite element simulation using the High Frequency Structure Simulator (HFSS) software, see SI for a detailed description. Figure 2(a) illustrates the spatial distribution of the electric field ($E$), corresponding to the TE101 mode of the 3D cavity without the device chip inside. The inclusion of the recess does not significantly alter the distribution of $E$. Figure 2(c) shows the distribution of $E$ along the central axis of the recess (the $y$-axis labeled in Fig. 2(a)). Upon inserting a device chip, the superconducting antenna behaves as an electric dipole and significantly modifies the distribution of $E$ \cite{antenna_1, antenna_2}. In the simulation, we have simplified the nanowire region as a lumped inductor (20 nH). Figure 2(b) presents an overview of this distribution (TE101), while Figure 2(d) shows a line cut along the $y$-axis. The left peak in Fig. 2(d) corresponds to the edge of the antenna strip, where $E$ tends to be the strongest near the sharp edges of a conductor, as shown in the inset of Fig. 2(b). The recess height is designed large enough to minimize the proximity effect from the antenna to the recess (the left peak in Fig. 2(d)), see Fig. S1 for the simulation. For the field distributions of other microwave modes and the qubit mode, see Fig. S2 in SI.

We then simulated the coupling strength between the SMA connector and the cavity mode (TE101), represented by the coupling quality factor $Q_c$. $Q_c$ was extracted by fitting the simulated reflection coefficient of the cavity in the driven mode, using the formula in Ref.\cite{Gross_2022}. Figure 2(e) depicts $Q_c$ as a function of the length $d$ of the SMA pin inserted into the cavity (without the device chip). As the SMA pin penetrates deeper into the cavity, $Q_c$ decreases, indicating a stronger coupling between the cavity and the probe. Figure 2(f) shows a similar trend when a device chip is inserted. The red dots in the panels of Fig. 2 correspond to the actual parameters of device A. The $Q_c$ value of 10 K in Fig. 2(f) is slightly higher that in Fig. 2(e) (7 K), possibly due to the presence of the chip substrate modifying the field distribution of the cavity mode. We have simulated a test case with only the device substrate (without the antenna and nanowire devices), the calculated $Q_c$ is also around 10 K, suggesting the significant role of the substrate.

Figure 2(g) shows the qubit-cavity coupling strength, $g$, as a function of $L$, the length of part of the antenna that is inside the cavity (see Fig. 1(e) for its labeling). The part of the antenna located inside the recess is 1.55 mm in length and kept fixed.  To extract $g$, we used the method of energy participation ratio to calculate the cross-Kerr coefficient which is a function of $g$ \cite{EPR, RMP_cQED}, see SI for details. Increasing $L$ results in a stronger $g$ and a smaller charging energy $E_C$ of the qubit, as shown in Fig. 2(h). An $E_C$ of $\sim$ 190 MHz (the red dot) is typical for a gatemon qubit \cite{Huo_gatemon}. In Figs. 2(i-j), we varied the antenna width ($w$) and simulated the changes in $g$ and $E_C$, respectively. The values of the red dots slightly differ between Figs. 2(g-h) and 2(i-j), likely due to the different mesh shapes used in the finite element analysis.

Next, we characterize the reflection coefficient of the 3D cavity, loaded into a dilution fridge with a base temperature below 50 mK. Note that the labeling , $S_{21}$, refers to the vector network analyzer, while for the cavity, $S_{11}$ (reflection coefficient)  was measured throughout this work, see Fig. S3 in SI for circuit details on input and output connections. Figure 3(a) shows $|S_{21}|$ without a device chip inside the cavity. Prior to measurement, the cavity was annealed in dry air \cite{Copper_anneal} to enhance its quality factor. Figure S4 in SI shows its post-annealing photograph (Figure 1(a) is the one before annealing). The microwave probe power at the cavity port 1 was about -96 dBm, calculated based on the VNA output power and the circuit attenuators (Fig. S3). If the losses of the microwave cables were taken into account ($\sim$ -20 dBm attenuation at 5 GHz), the actual power at the cavity port 1 would be $\sim$ -120 dBm. The cavity resonant frequency is $f_{\text{C}}$ = 5.443 GHz. The blue curve is fitted based on the formula $S_{21}=A \Big [1-2\frac{Q_l}{|Q_c\text{cos}(\theta)|}e^{i\theta}/(1+2iQ_l\frac{f_{\text{r}}-f_{\text{C}}}{f_{\text{C}}}) \Big ]$ \cite{Gross_2022}. The coupling quality factor $Q_c$ is estimated to be $\sim$ 7360, consistent with the simulation in Fig. 2(e). The loaded (total) quality factor, $Q_l=(1/Q_i+1/Q_c)^{-1}$, is $\sim$ 6740. We infer the internal quality factor $Q_{i}$ to be (80 $\pm$ 5) $\times$ $10^3$.  Figure 3(b) shows $S_{21}$ in the complex plane where the fitting agrees reasonably well with the experimental data.  

\begin{figure}[ht]
\includegraphics[width=\columnwidth]{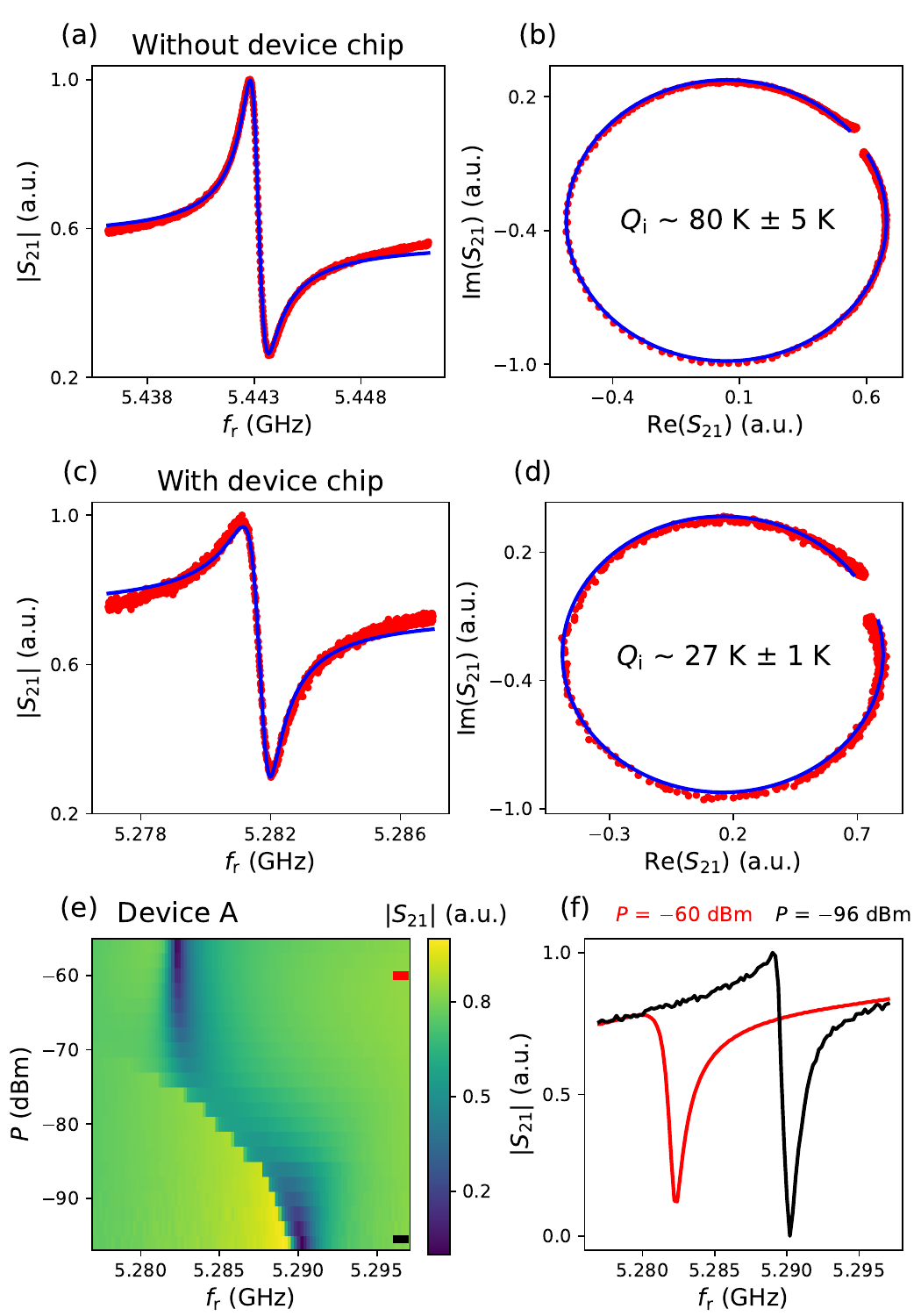}
\centering
\caption{Cavity reflection. (a) $|S_{21}|$ as a function of probe frequency $f_{\text{r}}$. The blue line is a fit. The cavity has no device chip inside. (b) $S_{21}$ vs $f_{\text{r}}$ in the complex plane. (c-d) Reflection coefficient of the same cavity with a device chip (device A) inserted. $V_{\text{G}}$ = -8 V. (e) $|S_{21}|$ vs $f_{\text{r}}$ and the probe power $P$ for device A. $V_{\text{G}}$ = 11.0 V. (f) Line cuts from (e) at the high power (red) and the low power (black) regimes. }
\label{fig3}
\end{figure}

Figures 3(c-d) show the $S_{21}$ measurement of the same cavity, but with a chip (device A) inserted. The probe power was about -96 dBm, the same as in the no-chip case. The gate voltage $V_{\text{G}}$ was set to -8.0 V, pinching off the Josephson junction and making the qubit frequency far away from $f_{\text{C}}$. The fitting (blue curve) yields an internal quality factor $Q_i \sim$ (27 $\pm$ 1) $\times$ $10^3$, a decent value, although significantly smaller than the no-chip case. This reduction in $Q_i$ is attributed to the loss caused by the device chip. The resonant frequency $f_{\text{C}}$ = 5.2816 GHz differs from the no-chip case by roughly 160 MHz. This difference is likely due to the redistribution of the electric field because of the device chip (and the antenna). The simulated $f_{\text{C}}$ in Figs. 2(a) and 2(b) are 5.44 GHz and 5.24 GHz, roughly consistent with the experimental data in Figs. 3(a-d). $Q_c$ is extracted to be $\sim$ 7270, slightly lower than the simulation in Fig. 2(f). This discrepancy likely arises from minor variation of the penetration length of the SMA pin upon reloading. For the power dependence of $Q_i$, see Fig. S4 in SI.

\begin{figure*}[ht]
\includegraphics[width=0.94\textwidth]{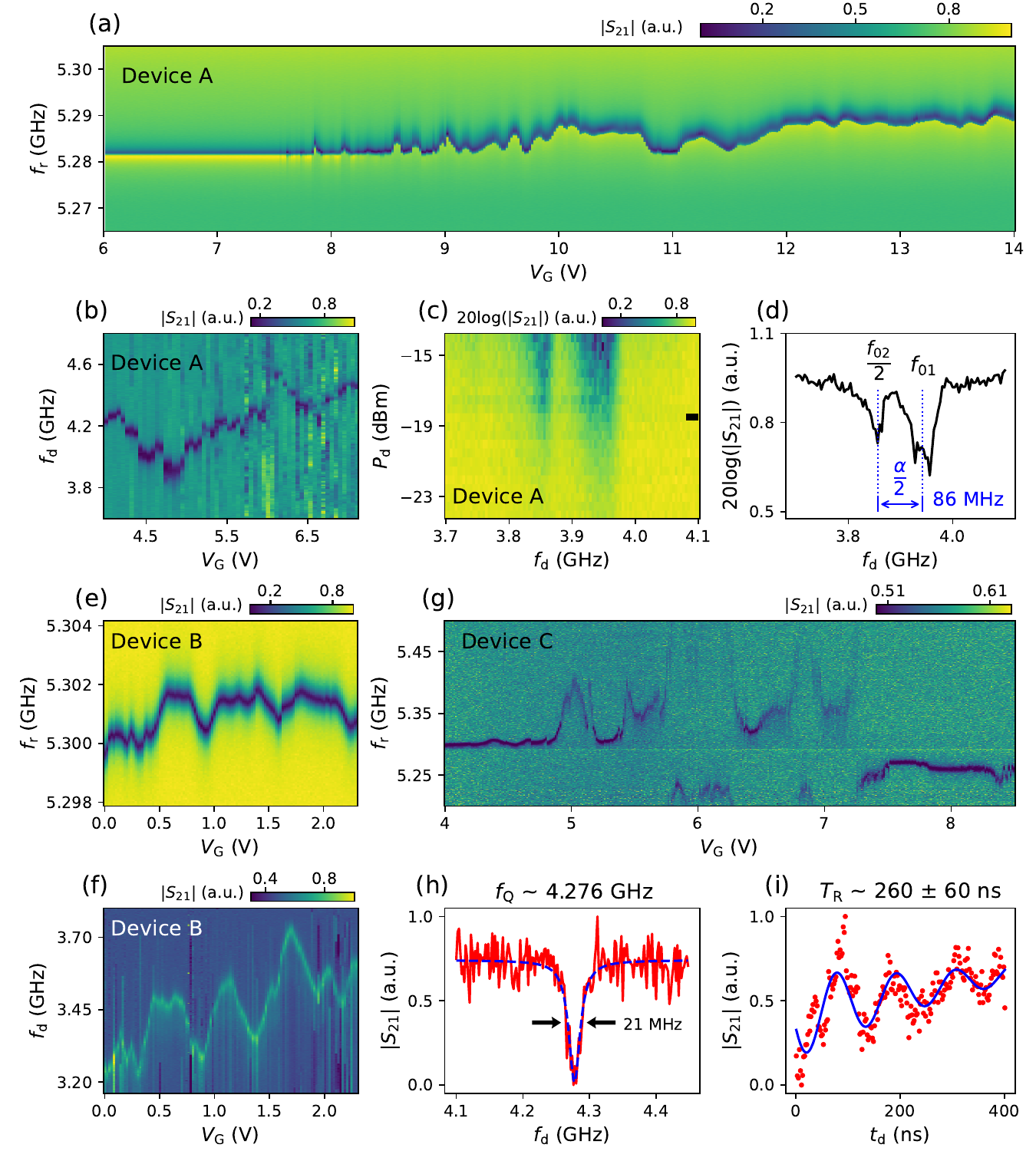}
\centering
\caption{Gate-tunable cavity shift and two-tone spectroscopy. (a) Gate dependence of the cavity shift of device A. (b) Two-tone spectroscopy of device A. (c) Spectroscopy as a function of qubit drive power ($P_{\text{d}}$). (d) Line cut from (c) (see the black bar), resolving the two-photon transition. (e) Gate dependence of the cavity shift of device B. (f) Two-tone spectroscopy of device B. (g) Gate dependence of the cavity shift of device C. (h) Two-tone spectroscopy of device C. The blue line is a Lorentzian fit. (i) Rabi oscillation in time-domain measurement.  $V_{\text{G}}$ = 3.88 V for (h-i).}
\label{fig4}
\end{figure*}

We then set $V_{\text{G}}$ to 11.0 V to activate the Josephson element which brought the qubit frequency close to $f_{\text{C}}$. In this regime, the interaction between the qubit and the cavity can be observed in the power ($P$) dependence of the cavity reflection, as shown in Fig. 3(e). The shift of the cavity resonant frequency from the high probe power regime (the red curve in Fig. 3(f)) to the lower probe power regime (the black curve) is the cavity-qubit dispersive shift. A lower $Q_i$ of the cavity ($\sim$ 8300) is extracted due to the qubit-induced cavity relaxation (the Purcell effect) \cite{Purcell}.

Figure 4(a) shows the gate dependence of the cavity shift for device A at the power of -86 dBm. At low $V_{\text{G}}$, the Josephson element is pinched off, and the bare cavity frequency ($f_{\text{bare}}$) is resolved. As $V_{\text{G}}$ increases, the Josephson element is turned on, causing the cavity shift. The shift amplitude $\chi/2\pi \equiv f_{\text{C}}-f_{\text{bare}}$ can be calculated (in the dispersive regime) to be $g^2/\Delta$ \cite{Koch_PRA}, where the detuning $\Delta/2\pi = f_{\text{bare}}-f_{\text{Q}}$. $f_{\text{Q}}$ is the qubit frequency. As $hf_{\text{Q}} = E_{01} \sim \sqrt{8E_J E_C}$, and $E_{J} (I_c)$ is a function of $V_{\text{G}}$ (the critical current $I_c$ is gate-tunable), the cavity shift is also $V_{\text{G}}$ dependent. The fluctuations in Fig. 4(a) are due to a nonmonotonic dependence of $I_c$ on $V_{\text{G}}$, which is typical for nanowire Josephson junctions. An ``anti-crossing'' feature is not observed for this device, suggesting that the maximum of $f_{\text{Q}}$ does not exceed $f_{\text{C}}$, due to the limited gate-tunability of $I_c$.

We then carried out two-tone spectroscopy by fixing the readout frequency near $f_{\text{C}}$ and scanning the qubit drive frequency $f_{\text{d}}$, as shown in Fig. 4(b). The dark dip denotes the qubit resonance/energy which is $V_{\text{G}}$ dependent. The range of $V_{\text{G}}$ in Fig. 4(b) does not match that in Fig. 4(a) due to the gate hysteresis, see Fig. S5 in SI. $g=\sqrt{\chi\Delta}$ could also be estimated from Fig. 4(b) as both $f_C$ and $f_Q$ can be extracted, see Fig. S6 in SI for details. The estimated $g/2\pi \sim 100$ MHz is consistent with the simulation in Figs. 2(g) and 2(i).

Figure 4(c) shows the qubit spectrum as a function of the drive power $P_{\text{d}}$. A second dip at a lower energy appears, see Fig. 4(d) for a line cut. This dip is a two-photon process, corresponding to a transition from $\ket{0}$ to $\ket{2}$ (the second excited state). Its energy is thus $f_{02}/2$ while the qubit energy $f_{01}=E_{01}/h$. From the spacing of the two dips $f_{02}/2-f_{01}=\alpha/2$, we can infer the qubit anharmonicity $\alpha = f_{12}-f_{01}$ $\sim$ -172 MHz, roughly matching the simulated charging energy (the red dot in Fig. 2). Unlike transmons, the anharmonicity for a gatemon qubit may not be $-E_C$ but between $-E_C$ and $-E_C/4$, depending on the transmission probability of the Andreev modes in the Josephson junction \cite{Anharmonicity},  see SI for an estimation of this probability. Given that the Andreev modes are gate-tunable, the anharmonicity is not constant but also gate dependent.

In Figures 4(e-f), we show the single-tone and two-tone measurement of a second device (device B). The gate dependence of the qubit spectroscopy in Fig. 4(f) roughly matches with the cavity shift in Fig. 4(e). The deviations at e.g. $\sim$ 1.2 V and 1.8 V are likely due to the mesoscopic instabilities in the device. Figure 4(g) exhibits a third device (device C), where anti-crossings can be revealed, see e.g. $V_{\text{G}}$ $\sim$ 5.8 V, 6.3 V and 7 V. These anti-crossings suggest that the qubit frequency can be tuned to match and exceed the cavity frequency.

We then set $V_{\text{G}}$ at 3.88 V and performed the two-tone spectroscopy for device C, see the red curve in Fig. 4(h). The blue line is a Lorentzian fit of the qubit line-shape. The full width at half maximum (FWHM) is $\sim$ 21 MHz. Figure 4(i) shows the time domain measurement of device C by varying the duration time ($t_{\text{d}}$) of the qubit drive. Rabi oscillations were observed (red dots). The blue line is a fit using the formula: $y=A \cdot \text{exp}(-t_{\text{d}}/T_{\text{R}})\cdot \text{cos}(\omega t_{\text{d}}+B) +at_{\text{d}}+b$. From this fit, we extract a Rabi coherence time $T_{\text{R}}$ = 260 $\pm$ 60 ns. The upward slope of the background, also observed in Ref.\cite{CPH_Pi_qubit, SAGmon}, likely originates from the leakage to higher level states.  Further time-domain measurements for $T_1$ and $T_2^*$ are unsuccessful due to device instabilities, similar to devices A and B. The limiting factor for further time-domain manipulation of the qubit likely lies in the quality of the device, such as contacts and gates. The superconducting film (Nb) exhibits a poor quality with a low critical temperature ($T_c$) of 3.9 K. This $T_c$ is significantly lower than the typical value ($\sim$ 9 K), possibliy due to the low sputtering rate (13 nm/min). Future improvements on increasing this rate are necessary for higher $T_c$ and thinner Nb films. For additional data of devices A, B and C, we refer to Figs. S5 and S6 in SI.  In Figs. 4(b), 4(f) and 4(g), a background was subtracted to enhance visibility, see Fig. S7 for details. In addition to the copper cavity, we also conducted similar experiments using a 3D aluminum cavity with a similar design, and the corresponding results are presented in Fig. S8.

In summary, we have proposed and implemented a gate-compatible 3D cavity architecture for circuit QED experiments. By incorporating an InAs-Al nanowire Josephson device into a recess machined on the sidewall of the cavity, we achieved a cavity internal quality factor of 27 000. A long superconducting strip couples the device to the cavity mode and forms a gatemon qubit with the Josephson junction. Gate-tunable cavity shift and two-tone qubit spectroscopy have been demonstrated. Our architecture allows the probing of gate-tunable quantum devices in a 3D microwave cavity.  Future works could study the magnetic field compatibility, requiring thinner Nb films and higher film quality. Note that although the copper cavity should be magnetic field resilient for all field directions, the device cannot survive a large perpendicular field due to vortex formation in the antenna. Other circuit designs, e.g. reducing the antenna width, are needed if a large perpendicular field is required.

\textbf{Acknowledgment} We thank valuable discussions with Luyan Sun. This work was supported by Tsinghua University Initiative Scientific Research Program, the National Natural Science Foundation of China (Grants 61974138, 92065206, 92065106 and 12374459), the Innovation Program for Quantum Science and Technology (Grant No.2021ZD0302400) and Youth Innovation Promotion Association, Chinese Academy of Sciences (nos. 2017156 and Y2021043). Raw data and processing code within this paper are available at https://doi.org/10.5281/zenodo.10703169

\bibliography{mybibfile}% Produces the bibliography via BibTeX.

\newpage

\onecolumngrid

\newpage


\section*{Supplementary material}

# Supplemental Information for "Gate-Compatible Circuit Quantum Electrodynamics in a Three-Dimensional Cavity Architecture"

Zezhou Xia,[1, *] Jierong Huo,[1, *] Zonglin Li,[1, *] Jianhua Ying,[2, *] Yulong Liu,[3, *] Xin-Yi Tang,[1] Yuqing Wang,[3] Mo Chen,[3] Dong Pan,[4] Shan Zhang,[1] Qichun Liu,[3] Tiefu Li,[5, 3] Lin Li,[3] Ke He,[1, 3, 6, 7] Jianhua Zhao,[4] Runan Shang,[3, 7] and Hao Zhang[1, 3, 6, †]

[1]State Key Laboratory of Low Dimensional Quantum Physics,
Department of Physics, Tsinghua University, Beijing 100084, China
[2]Yangtze Delta Region Industrial Innovation Center of Quantum and Information, Suzhou 215133, China
[3]Beijing Academy of Quantum Information Sciences, Beijing 100193, China
[4]State Key Laboratory of Superlattices and Microstructures, Institute of Semiconductors,
Chinese Academy of Sciences, P. O. Box 912, Beijing 100083, China
[5]School of Integrated Circuits and Frontier Science Center for
Quantum Information, Tsinghua University, Beijing 100084, China
[6]Frontier Science Center for Quantum Information, Beijing 100084, China
[7]Hefei National Laboratory, Hefei 230088, China

## Device fabrication and simulation

### Antenna fabrication

A 100-nm-thick Nb superconducting film was first sputtered onto a high-resistance-silicon substrate. The sputtering pressure was 5 mTorr in an argon environment. The photoresist, S1813, was spun onto the Nb film (3000 rpm, 60 s) and baked at 115 °C for 120 s. Then direct laser writing was used to define the patterns for the antenna, markers and gate line. After developing in AZ for 1 min, reaction ion etching ($O_2$ pressure 5 Pa, 50 W for 20 s, $CF_4$ pressure 2 Pa, 100 W for 165 s) was performed to etch away the Nb film region uncovered by the resist. Finally, the residual resist was dissolved in acetone.

### Nanowire device fabrication

Thin InAs nanowires were grown by molecular beam epitaxy, followed by an in-situ deposition of an Al film (half shell, 15 nm thick). These hybrid wires were then transferred, through wiping of clean room tissues, from the growth chip onto the Si substrate where an antenna has been fabricated. PMMA 672.045 (A4.5) resist was then spun at 4000 rpm for 1 min and baked at 120 °C for 10 min. Electron beam lithography (EBL) was performed to pattern the etch windows. After development in MIBK:IPA = 1:3 for 50 s and post-baking at 130 °C for 3 min, the chip was immersed in Transene Aluminum Etchant Type D at 50 °C for 10 s. After the wet etching, the resist was removed in acetone. Another EBL (resist PMMA 671.05) was then performed for the contacts and side gate electrodes by sputtering Ti and Nb (thickness 1 nm and 100 nm) for devices B and C; for device A, it was Ti/NbN (1 nm/100 nm); for device D, it was Ti/NbTiN (1 nm/100 nm). Before the sputtering, a short argon plasma etching (90 s, 50 W, 0.05 Torr) was performed in situ to ensure good ohmic contact.

### Finite element simulation

To solve the eigenmode, the copper conductivity of the cavity was assumed to be $1.5 \times 10^{10}$ $(\Omega \cdot m)^{-1}$. The SMA pin was assigned to be copper beryllium alloy (conductivity $1.55 \times 10^7$ $(\Omega \cdot m)^{-1}$) from the database. The dielectric constant of the substrate (Si) was set to be 11.9 with a dielectric loss tangent of $1.5 \times 10^{-7}$.

We assumed the antenna film as a 2D conductor with a perfect electrical boundary, i.e. $E$ is perpendicular to its surface (a perfect conductor). We made this assumption because the resistance of a superconducting film is negligible, and the thickness of 100 nm is thin enough compared to the antenna size and the wavelength of microwaves. The InAs-Al Josephson junction was simplified as an inductor with an inductance of 20 nH, as the nanowire in a gatemon is a Josephson junction which can be treated as an inductor to the first order approximation.

The geometric size of the contacts near the device was enlarged to be compatible with the mesh size and to reduce computational cost. $Q_c$ was extracted by fitting the simulated reflection coefficient near the cavity resonant frequency. To calculate the qubit parameters, the rounded corners of the cavity was assumed to have rectangular shapes in the model and the cavity was set to be a perfect conductor. $E_C$ was extracted from the calculated qubit frequency $hf_Q = \sqrt{8E_J E_C}$.

HFSS simulations typically do not provide direct information about $g$. To extract $g$, we used the method of energy participation ratio (EPR) to calculate the cross-Kerr coefficient via the python module pyEPR (pyEPR-quantum), following Ref. [54] (the cross-Kerr coefficient is a function of $g$). We first expend and separate the Josephson energy

---

* equal contribution
† hzquantum@mail.tsinghua.edu.cn

term $(-E_J \cos\varphi_J)$ in the Hamiltonian into the linear and non-linear parts: $H = H_{lin} + H_{nl} = \hbar\omega_C a_C^\dagger a_C + \hbar\omega_Q a_Q^\dagger a_Q - E_J[1 + \frac{1}{24}\varphi_J^4 - \frac{1}{720}\varphi_J^6 + ...]$. $\varphi_J = \varphi_C(a_C^\dagger + a_C) + \varphi_Q(a_Q^\dagger + a_Q)$. $\varphi_C$ and $\varphi_Q$ are the quantum zero-field fluctuations of the junction flux in the cavity and qubit mode, respectively. We then calculate the EPR ($p_m$) of the junction in the mode $m$ ($m$ = cavity or qubit), defined to be the fraction of inductive energy stored in the junction relative to the total inductive energy stored in the entire circuit. $p_m$ is proportional to $\varphi_C^2$ and $\varphi_Q^2$, and can also be calculated using the EM field distribution by definition. Note that the total electrical energy and total inductive energy of a resonant system are equal. Thus, we can calculate the quantum Hamiltonian based on the classical EM field. From $p_m = \frac{\langle\psi_m|1/2E_J\varphi_J^2|\psi_m\rangle}{\langle\psi_m|1/2H_{lin}|\psi_m\rangle}$, we can get $\varphi_C^2 = \frac{p_C\hbar\omega_C}{2E_J}$, $\varphi_Q^2 = \frac{p_Q\hbar\omega_Q}{2E_J}$. We can simulate $p_m = \frac{E_{jj}}{E_{ind/tot}} = \frac{E_{ind/tot} - E_{ind/field}}{E_{ind/tot}} = \frac{E_{cap/field} - E_{ind/field}}{E_{cap/field}}$. We can then get the cross-Kerr coefficient $\chi_{QC} = -2\frac{g^2 E_C/\hbar}{(\omega_Q - \omega_C)(\omega_Q - \omega_C - E_C/\hbar)}$, from which $g$ can be estimated.

**Estimation of the Josephson junction transparency**

The anharmonicity $\alpha = -E_C(1 - \frac{3}{4}\frac{\sum T_i^2}{\sum T_i})$, $T_i$ is the transparency of the $i$<sup>th</sup> occupied Andreev mode. For simplicity, we assume N modes with equal transparency $T = \frac{\sum T_i}{N}$. In Fig. 4(d), $\alpha \sim -172$ MHz. From the simulation, we get $E_C \sim$178-192 MHz (take the average 185 MHz). $E_J = (hf_Q)^2/8E_c \sim 10$ GHz. And $E_J = \frac{\Delta}{4}\sum T_i$, where $\Delta$ is the superconducting gap, $\sim 300$ µeV (75 GHz) based on our transport measurement. Therefore, $\sum T_i \sim 0.533$. $\alpha = -E_C(1 - \frac{3}{4}\frac{\sum T_i}{N})$, we get $N \sim 6$ and $T \sim 0.09$.

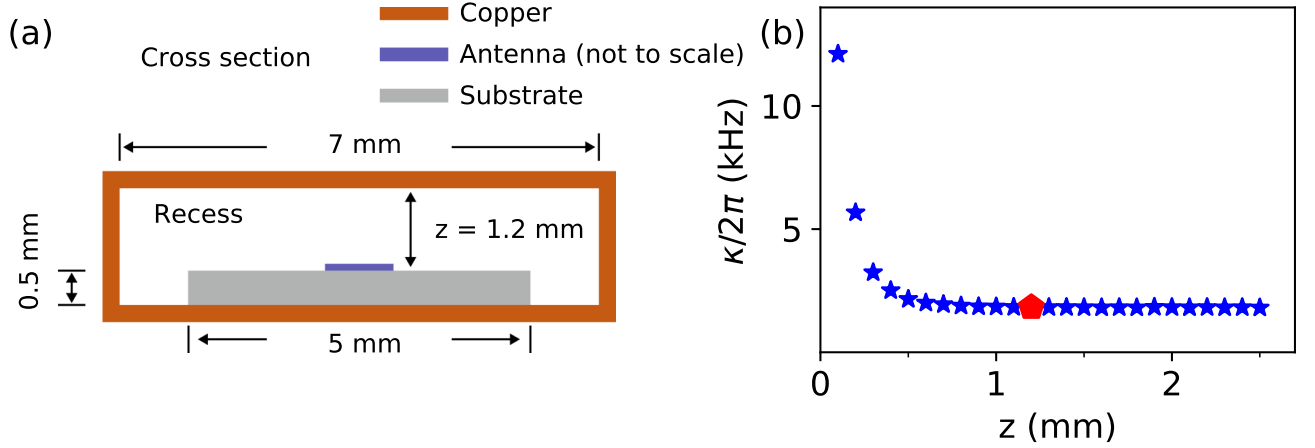

FIG. S1. (a) Schematic of the recess cross section. The antenna (100 nm thick and 0.2 mm wide) is not drawn to scale for visibility. The spacing between the antenna and the recess top wall is $z = 1.2$ mm. (b) Simulated dissipation-induced broadening of the qubit energy, $\kappa/2\pi$, as a function of $z$. As $z$ approaches zero, the qubit broadening increases due to ohimc dissipation in the recess side wall. This dissipation is caused by the proximity of the qubit electric field from the antenna to the copper recess. The current design (red dot) is in the saturated regime, suggesting that this proximity effect is not a limiting factor for the qubit coherence.

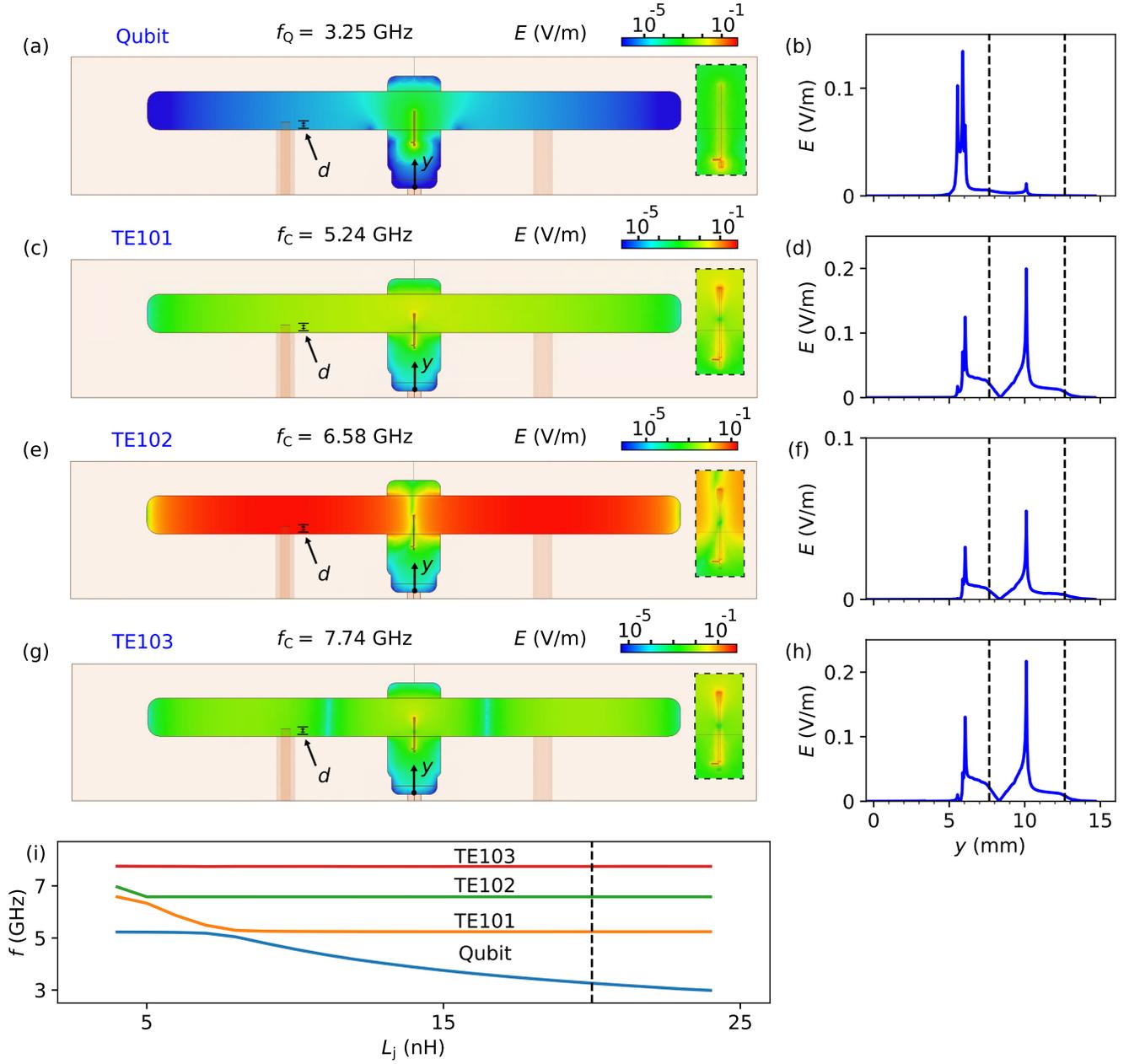

FIG. S2. Simulated electric field distribution of qubit and cavity modes. (a) Qubit mode with a frequency of 3.25 GHz. (b) Line cut along the $y$ axis. (c-d) TE101 mode, replot of Figs. 2(b) and 2(d). (e-h) TE102 and TE103 modes, with frequencies of 6.58 GHz and 7.74 GHz, respectively. (i) Frequencies of the qubit mode and the three cavity modes as a function of the nanowire Josephson inductance, $L_j$. Only the qubit frequency varies while the cavity frequencies are constants except at the anti-crossing points. The dashed line corresponds to the simulations in (a-h).

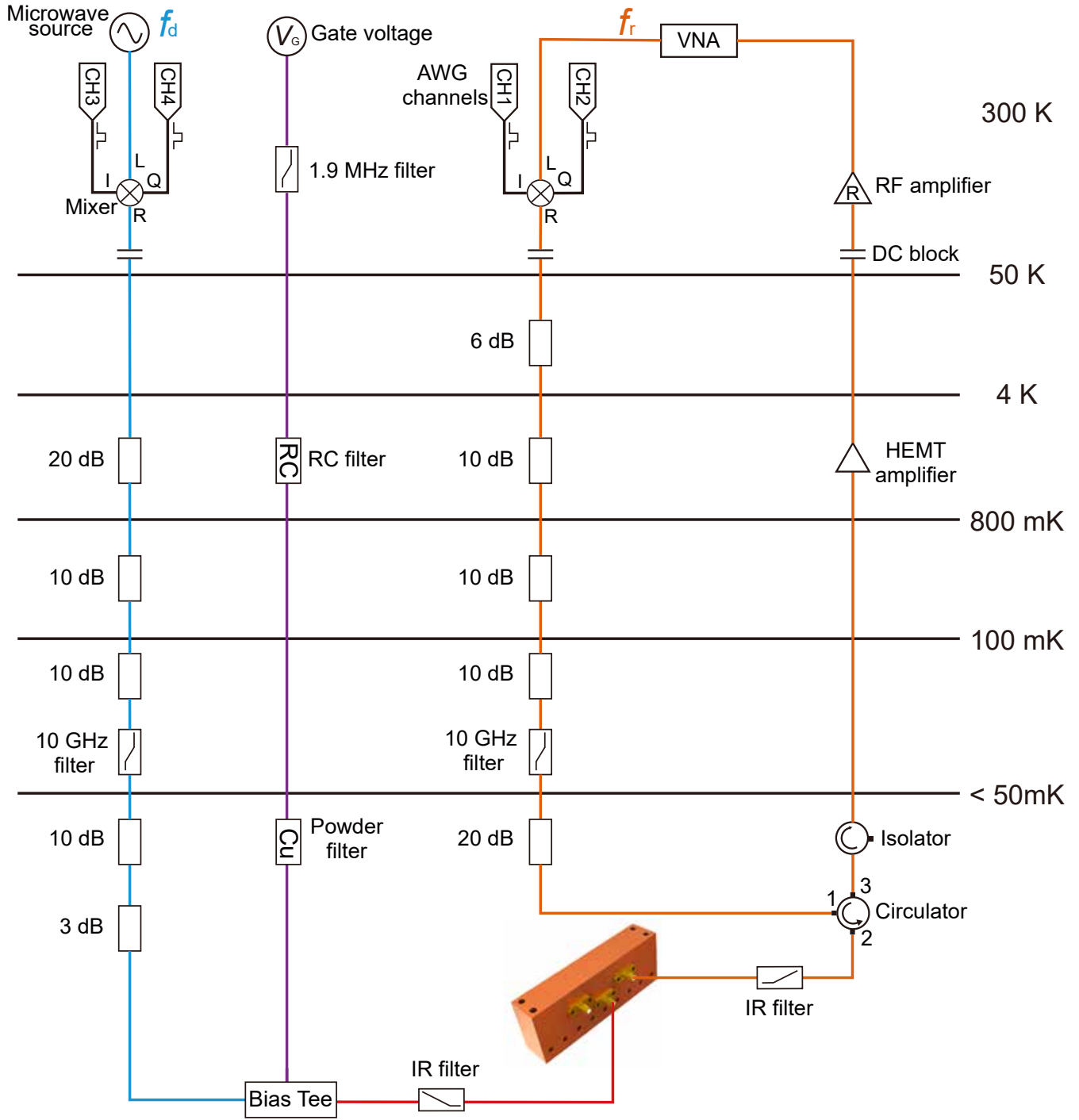

FIG. S3. Schematic of the measurement circuit. The purple line is used to apply a DC gate voltage to the device after passing through the RC filter and copper powder filter. The blue line is for the qubit control. A continuous microwave signal with frequency $f_d$ was generated by the microwave source. This signal was then modulated by DRAG wave pulses from an arbitrary waveform generator (AWG) through the mixer. The pulsed signal, after passing through several attenuators, was applied to the device gate line via a Bias Tee. The orange line is for cavity readout. A microwave signal (frequency $f_r$) generated by a vector network analyzer (VNA) was modulated by square wave pulses and fed to the SMA port of the cavity after passing through several attenuators. The reflected signal was collected through a circulator, amplified and then measured using the VNA.

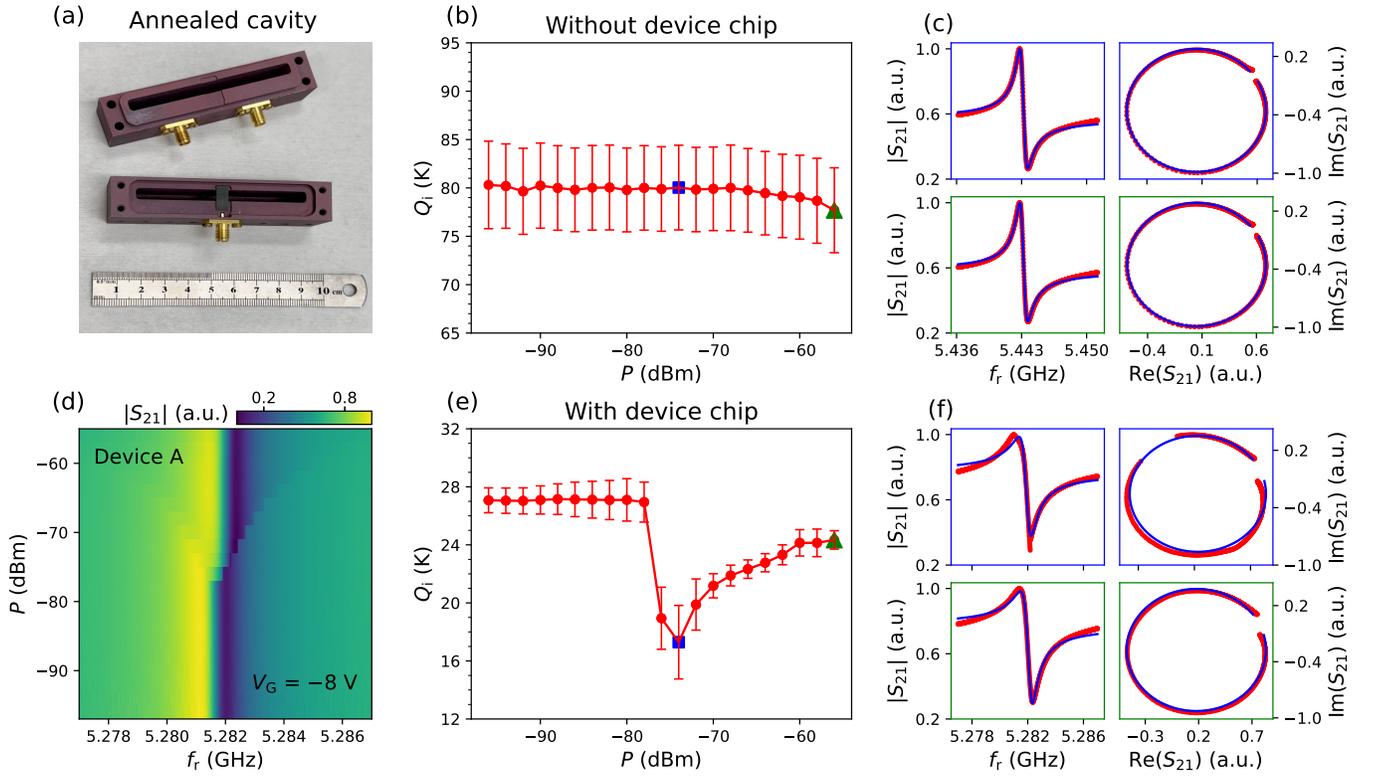

FIG. S4. (a) Post-annealing photograph of the 3D cavity. (b) Power dependence of $Q_i$ of the 3D cavity (after annealing) without the device chip inside. (c) Fitting of the cavity reflection at the power near -75 dBm and -55 dBm, respectively. (d) Power dependence of the cavity reflection with device A inside. $V_G$ = -8 V. (e) Extracted $Q_i$ from (d). (f) Two examples of the fitting at the power near -75 dBm and -55 dBm, respectively. The dip near -75 dBm in (e) is probably due to the Purell effect caused by a two-level system (see panel d), whose origin is currently unknown.

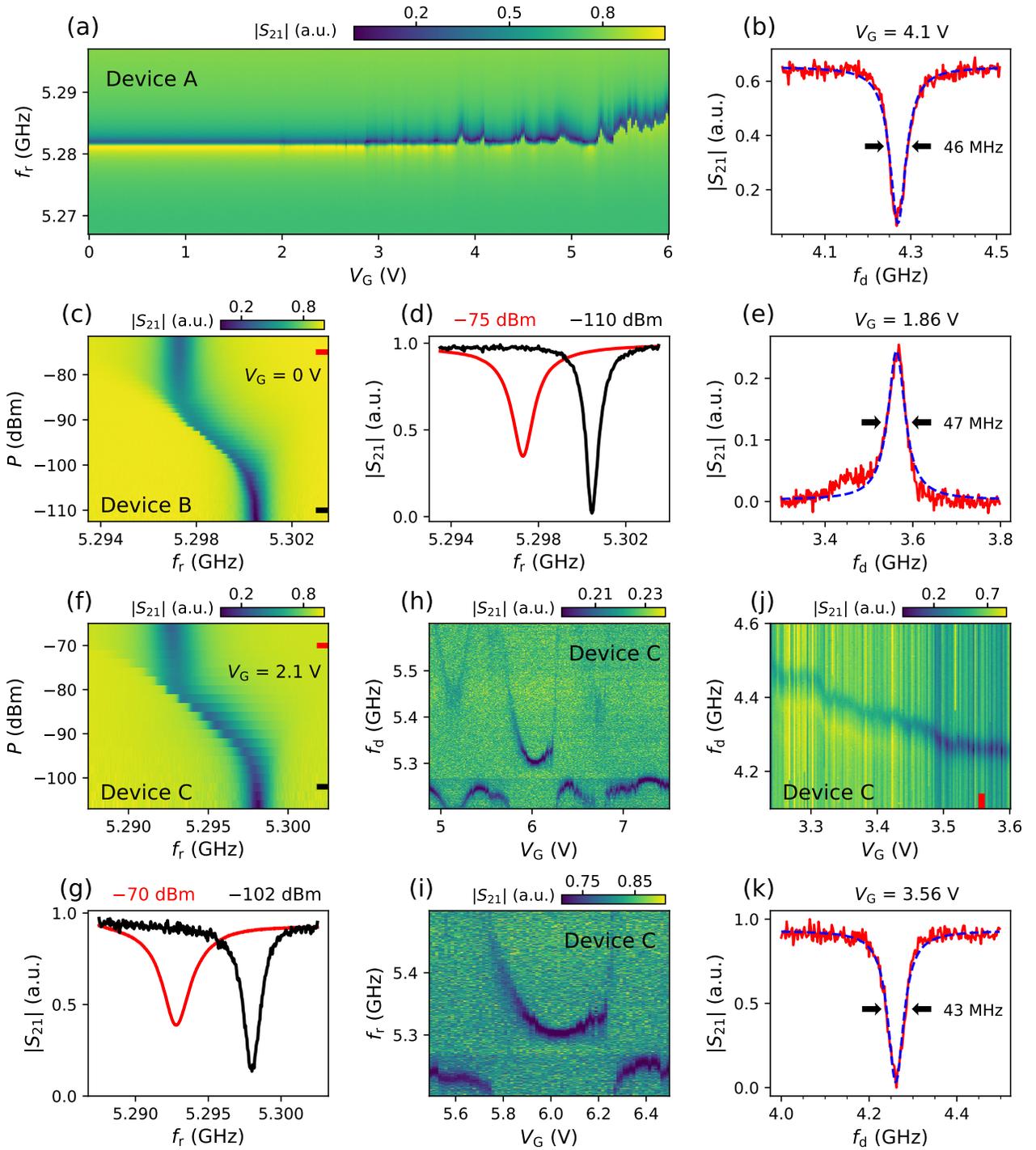

FIG. S5. (a) Gate dependence of the cavity shift of device A. The sweeping direction of $V_G$ was from 0 V to 6 V, the opposite of that in Fig. 4(a). (b) Two-tone spectroscopy, corresponding to a line cut in Fig. 4(b) at $V_G = 4.1$ V. The blue dashed line is a Lorentzian fit with a FWHM of 46 MHz. (c) Power dependence of the cavity reflection for device B. (d) Two line cuts of (c) at the high power (red) and low power (black) regimes, showing the cavity shift. (e) Two-tone spectroscopy of device B, corresponding to a line cut in Fig. 4(f) at $V_G = 1.86$ V. (f) Power dependence of the cavity refection for device C with line cuts shown in (g). (h) Gate dependence of the cavity reflection for device C. (i) An enlargement of (h) on the anti-crossing region. (j) Two-tone spectroscopy of device C. (k) A line cut from (j) with a FWHM of 43 MHz.

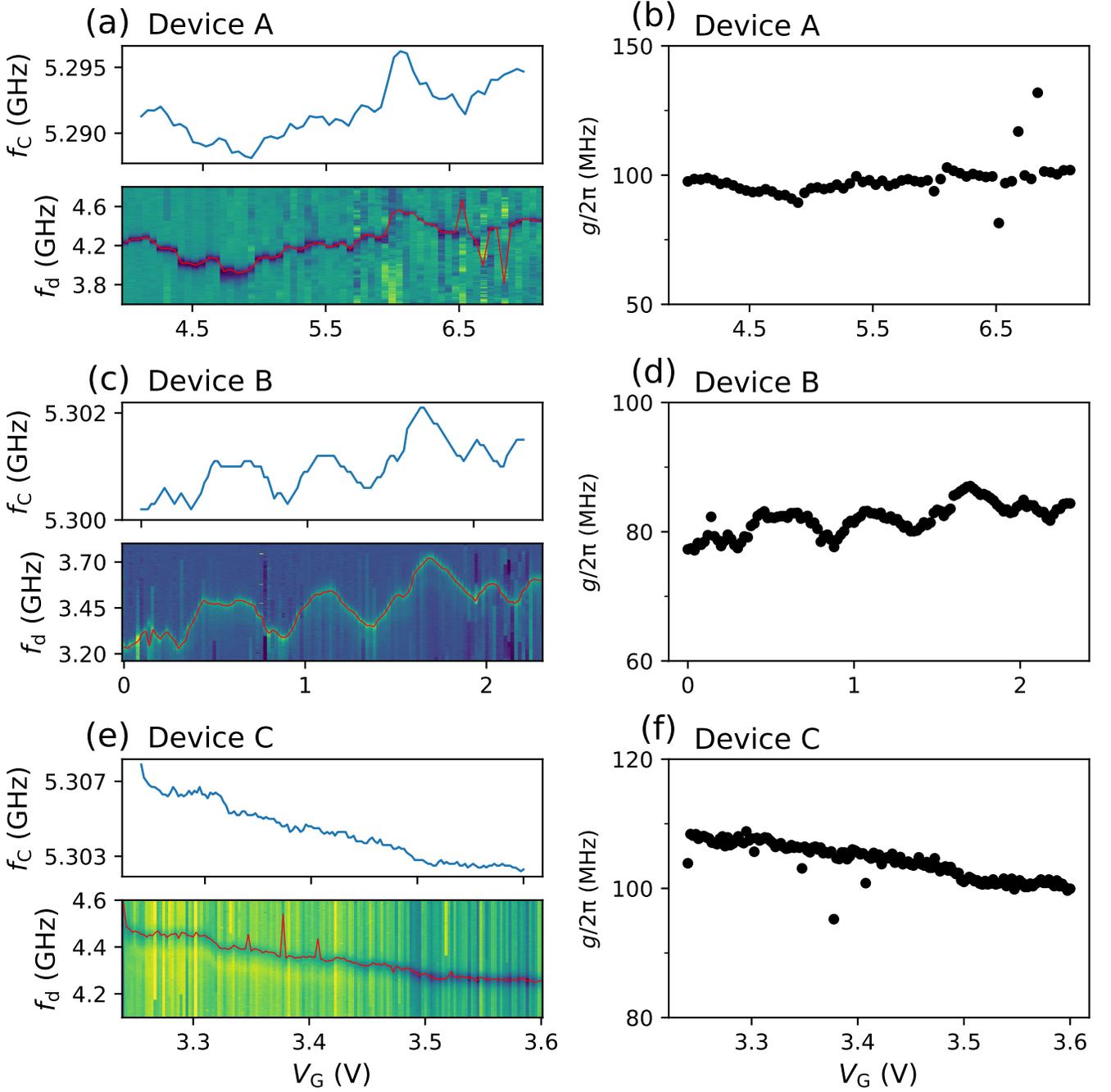

FIG. S6. Estimation of $g$. (a) Upper panel is the cavity resonant frequency $f_C$ as a function of $V_G$ for device A. Lower panel is the corresponding two-tone measurement (Fig. 4(b)) with the red line highlighting the qubit frequency $f_Q$. The actual measurement sequence was as follows: At each fixed $V_G$, before scanning the two-tone (qubit drive), the cavity drive was scanned first to extract $f_C$. We then calculate the dispersive shift $\chi/2\pi = f_C - f_{\text{bare}}$, and the detuning $\Delta/2\pi = f_{\text{bare}} - f_Q$. $g = \sqrt{\chi\Delta}$ can then be estimated, as shown in (b). $g \sim 100$ MHz is consistent with the simulation in Figs. 2(g) and 2(i). (c-f) Similar estimations for devices B and C. $g$ for device B ($\sim 80$ MHz) is slightly lower than the simulation, possibly due to the inaccurate placement of the chip.

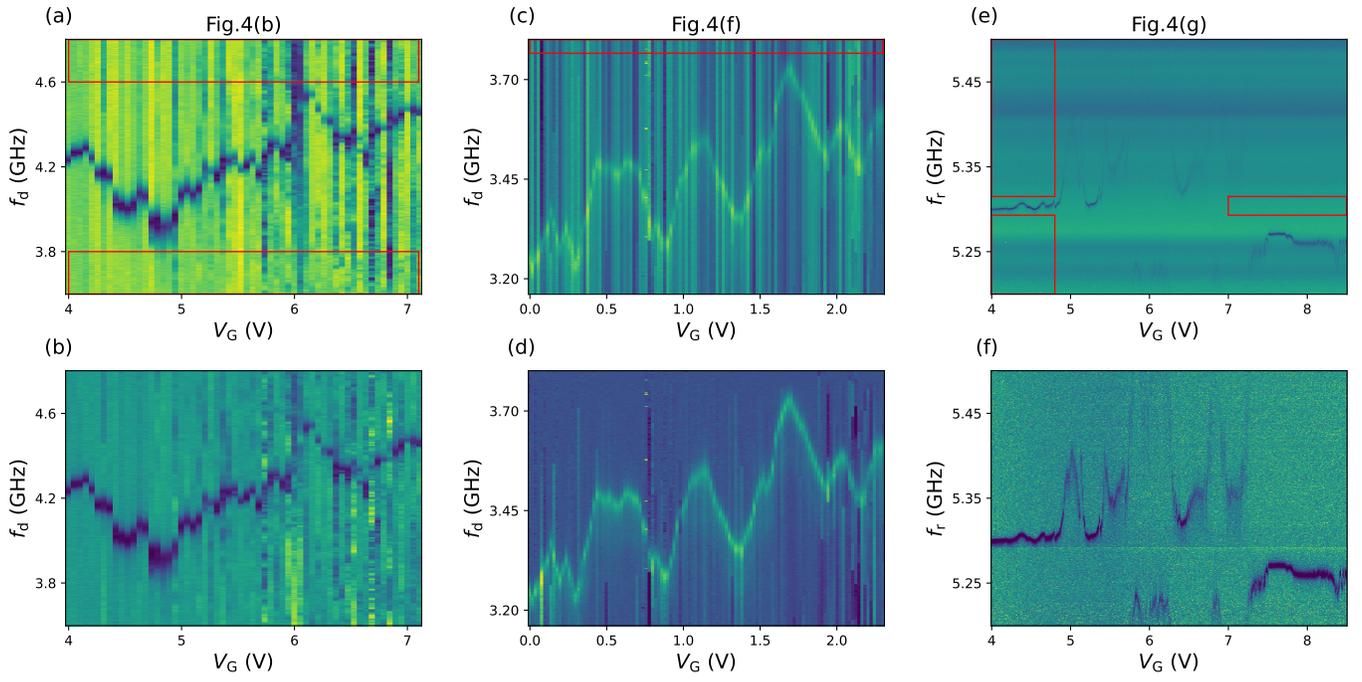

FIG. S7. Background subtraction. (a), (c) and (e) are the raw data corresponding to Figs. 4(b), 4(f) and 4(g), while the lower panels are the ones after the background subtraction. For (a) and (c), the signal within the red boxes was averaged for each value of $V_G$ as the background. The background was then subtracted from the raw data. For panel (e) the average was along the $V_G$ axis (another direction) within the red boxes.

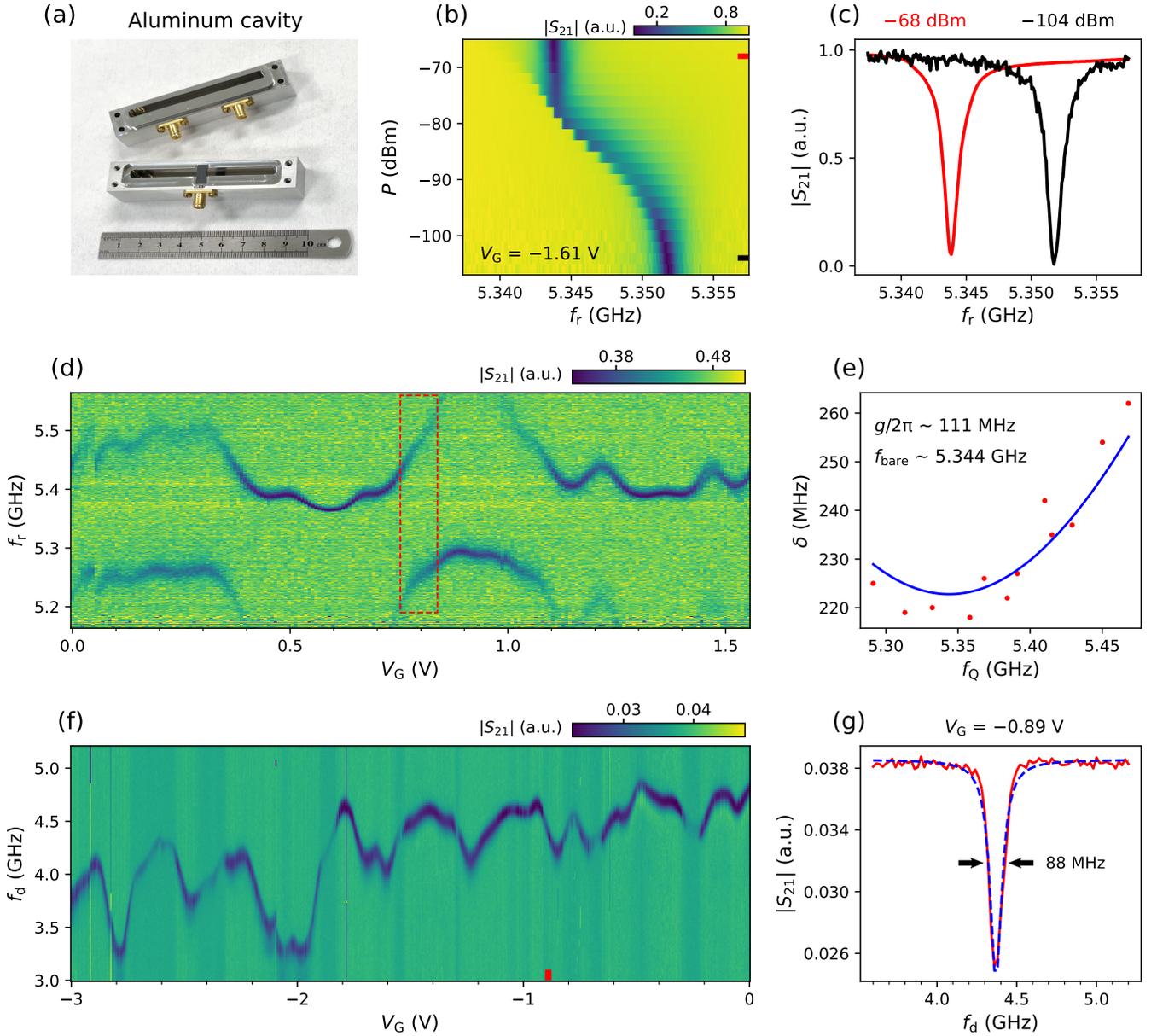

FIG. S8. Gatemon qubit in a 3D aluminum cavity. (a) Photograph of the 3D Al cavity with a device chip (device D). (b) Power dependence of the cavity shift. The reflection coefficient of the cavity was measured. (c) Line cuts from (b) at the high power (red) and low power (black) regimes. (d) Gate dependence of the cavity shift with clear anti-crossings. (e) The spacing $\delta = f_+ - f_-$ between the anti-crossing dips as a function of the qubit frequency ($f_Q = f_+ + f_- - f_{bare}$). $f_+$ and $f_-$ are the frequencies of the two dips within the red box in (d). The blue line is a fit based on the formula $\delta = \sqrt{(f_Q - f_{bare})^2 + 4(g/2\pi)^2}$ where $g/2\pi$ of 111 MHz can be extracted. (f) Two-tone qubit spectroscopy. (g) A line cut from (f) at $V_G = -0.89$ V. The blue dashed line is a Lorentzian fit.



\end{document}